\documentclass{iau307}
\usepackage{graphicx}
\usepackage{natbib}
\usepackage{url}
\usepackage{dtklogos}
\bibpunct{(}{)}{;}{a}{}{,}

\title[Short title] 
{The BinaMIcS project: understanding the origin of magnetic fields in massive stars through close binary systems}

\author[E. Alecian, C. Neiner, G.A. Wade and the BinaMIcS collaboration]   
{E. Alecian$^{1,2},$
  C. Neiner$^2$,
  G.A. Wade$^3$,
  S. Mathis$^{4,2}$,
  D. Bohlender$^5$,
  D. C\'ebron$^6$,
  C. Folsom$^7$,
  J. Grunhut$^8$,
  J.-B. Le Bouquin$^1$,
  V. Petit$^9$,
  H. Sana$^{10}$,
  A. Tkachenko$^{11}$,
  A. ud-Doula$^{12}$,
 \and the BinaMIcS collaboration }

\affiliation{$^1$UJF-Grenoble 1/CNRS-INSU, Institut de Plan\'etologie et d'Astrophysique de Grenoble (IPAG) UMR 5274, 38041 Grenoble, France,  email: {\tt evelyne.alecian@obs.ujf-grenoble.fr} \\[\affilskip]
$^2$LESIA, Observatoire de Paris, CNRS UMR 8109, UPMC, Universit\'e Paris Diderot, France\\[\affilskip]
$^3$Dept. of Physics, Royal Military College of Canada, Kingston, ON, Canada K7K 0C6\\[\affilskip]
$^4$Laboratoire AIM Paris-Saclay, CEA/DSM -- CNRS -- Universit\'e Paris Diderot, IRFU/SAp Centre de Saclay, 91191 Gif-sur-Yvette Cedex, France\\[\affilskip]
$^5$DAO, CNRC, Victoria, BC V9E 2E7, Canada\\[\affilskip]
$^6$Universit\'e Grenoble Alpes, CNRS, ISTerre, Grenoble, France\\[\affilskip]
$^7$LATT -- CNRS/Universit\'e de Toulouse, 14 Av. E. Belin, Toulouse F-31400, France\\[\affilskip]
$^8$ESO, Garching bei M\"unchen, Germany\\[\affilskip]
$^9$Bartol Research Institute, University of Delaware, Newark, DE 19716, USA\\[\affilskip]
$^{10}$ESA/Space Telescope Science Institute, Baltimore, MD 21218, USA\\[\affilskip]
$^{11}$Instituut voor Sterrenkunde, KU Leuven, Celestijnenlaan 200D, B-3001 Leuven, Belgium\\[\affilskip]
$^{12}$Penn State Worthington Scranton, Dunmore, PA 18512, USA\\[\affilskip]
}

\pubyear{2014}
\volume{307} 
\pagerange{}
\jname{New windows on massive stars: asteroseismology, interferometry, and spectropolarimetry}
\editors{G. Meynet, C. Georgy, J.H. Groh \& Ph. Stee, eds.}

\begin{document}

\maketitle

\begin{abstract}
It is now well established that a fraction of the massive ($M>8$~M$_{\odot}$) star population hosts strong, organised magnetic fields, most likely of fossil origin. The details of the generation and evolution of these fields are still poorly understood. The BinaMIcS project takes an important step towards the understanding of the interplay between binarity and magnetism during the stellar formation and evolution, and in particular the genesis of fossil fields, by studying the magnetic properties of close binary systems. The components of such systems are most likely formed together, at the same time and in the same environment, and can therefore help us to disentangle the role of initial conditions on the magnetic properties of the massive stars from other competing effects such as age or rotation. We present here the main scientific objectives of the BinaMIcS project, as well as preliminary results from the first year of observations from the associated ESPaDOnS and Narval spectropolarimetric surveys.
\keywords{stars: binaries: close, stars: magnetic fields}
\end{abstract}

\firstsection 
\section{Introduction}

The Binarity and Magnetic Interactions in various classes of stars (BinaMIcS) project proposes to bring new constraints on the physical processes of magnetic stars and of binary systems, by studying the interplay between magnetism and binarity. Binary stars have many advantages compared to single stars. Thanks to their gravitational interaction, model-independent measurements of the stellar fundamental properties (such as masses or radii) can be made. Binary systems also represent laboratories of novel physical processes through star-star interactions (e.g. tidal deformation and flows, mutual heating, wind-wind collisions, magnetospheric coupling, pulsation excitation). Finally, they bring the possibility to investigate the origin of magnetic fields in stars with similar formation and evolution scenarios, and the same age.

In the last decade, magnetic fields have been intensely studied in stars of various ages and masses. We now have a clear global picture of stellar magnetism: convective dynamos in the upper envelope of cool stars (below $\sim$6500 K) generate complex and variable magnetic fields, while in the hotter stars with radiative outer envelopes and masses above 1.5~M$_{\odot}$, fossil fields are observed in only 5-10\% of them \citep[see the review of][]{donati09,wade13}. Details on the origin and evolution of stellar magnetism of binary systems are not yet well understood. The BinaMIcS project\footnote{http://lesia.obspm.fr/BinaMIcS/} proposes to bring new constraints by addressing how star-star interaction in close binary systems can affect or be affected by magnetic fields. With this aim we have gathered a large international consortium of about 80 scientists expert in stellar magnetism and/or binary/multiple systems, observers, modellers, and theoreticians, and have proposed to address four scientific objectives.

\section{Scientific objectives}

\subsection{How do tidally-induced internal flows impact fossil or dynamo fields ?}

An initially eccentric binary system with non-synchronised, non-aligned components will tend to an asymptotic state with circular orbits, synchronised components, and aligned spins \citep[e.g.][]{hut80}. Such an evolution is possible thanks to a dissipation of the kinetic energy of internal flows into heat. During this process, three types of flows are generated in the stellar interiors: the equilibrium tide, a 3D-large scale flow induced by hydrostatic adjustment of a star in response to the perturbing gravitational field of the companion \citep[e.g.][]{remus12}; the dynamical tide corresponding to the excitation of eigen-modes of the stellar interior oscillations by the tidal potential \citep[e.g.][]{ogilvie07}; and the elliptical instability, a 3D-turbulent flow that can be excited in a rotating fluid of ellipsoidal shape \citep[e.g.][]{cebron14}. One of the objectives of the BinaMIcS project will be to address how such flows can affect the fossil fields of hot massive stars, or the dynamo fields in cool stars.

\subsection{How do magnetospheric Star-Star interactions modify stellar activity ?}

In hot or cool close systems, with two magnetic components, we expect important magnetic interactions between the components' magnetospheres. Intense magnetic reconnection phenomena may develop during the motion of the secondary (and its magnetosphere) through the magnetosphere of the primary, and vice-versa \citep[e.g.][]{gregory14}. We therefore aim at understanding the modification of stellar activity by such events.

\subsection{What is the magnetic impact on angular momentum exchanges ?}

In low-mass solar-type stars, stellar winds originate from the magnetic coronae \citep[e.g.][]{parker60}. In massive stars, winds are driven by spectral line absorption of the stellar radiation field \citep["line-driving",][]{castor75}. In both cases, these winds carry away angular momentum and slow down the rotation of the surface layers of the stars \citep[e.g.][]{uddoula09}. This phenomenon is amplified if the winds are magnetised \citep[e.g.][]{roxburgh83}. The combination of tidal friction and magnetised winds may have a significant impact on the evolution of binary systems \citep[e.g.][]{barker09}. We will bring observational constraints on the proposed impacts.

\subsection{What is the impact of magnetic fields on stellar formation, and vice-versa ?}

In hotter higher-mass stars ($M>1.5$~M$_{\odot}$), the fossil magnetic fields stored into their upper radiative zones are believed to be the results of fields accumulated or amplified during the star formation \citep[e.g.][]{borra82}. While this hypothesis explains well most of the properties of the fields of high-mass stars \citep{moss01}, one of the biggest challenges is to understand why only a small fraction (5-10\%) of high-mass stars host fossil fields \citep[e.g.][]{wade13}. A natural explanation would be the existence of fundamental differences in the initial conditions (IC) of star forming regions (e.g. local density, local magnetic strength ...). An excellent way to test this hypothesis is to study the magnetic fields in close binary systems, which contain two stars formed at the same time and with the same IC. Such a study will help us to disentangle IC effects from other (e.g. evolutionary) effects.

\section{Observing strategies and resources}

To address all the objectives exposed above, we focus on close binary systems (with orbital periods lower than 20 d), i.e. binary systems in which we expect a significant mutual interaction via tidal or magnetospheric interaction. The two components of such close binary systems are also very likely to have been formed from the same condensed molecular core, and therefore to share a similar history \citep[e.g.][]{bonnell94}, which is essential to study fossil field history.

We aim at acquiring high-resolution spectropolarimetric datasets of close binary systems of a sufficient quality to get sensitive measurements of the magnetic fields at the surface of both components. The origin of the surface magnetic fields in cool and hot stars being different, the observing strategy we have chosen is different. In the case of cool stars, magnetic fields are ubiquitous, and we have selected about 40 close binary systems with cool components in which indirect signs of magnetic activity ensure the detection of the magnetic fields in our data. Each system will be monitored and a magnetic map at the surface of the two components will be obtained.

In the case of hotter higher-mass binary systems, fossil fields are only expected in a small fraction of them. We have therefore built two samples. The first one, the Survey Component (SC), gathers about 220 close ($P_{\rm orb}< 20$~d) binary systems of magnitude V brighter than 8 mag in which we will look for magnetic fields (see the primary spectral types distribution in Fig. \ref{fig:spt}). We used three different catalogues to built this sample: the 9th catalogue of spectroscopic binary orbits \citep[SB9][]{pourbaix09}, the CHARA catalogue \citep{taylor03}, and a catalogue of O-type spectroscopic binaries \citep{sana11}. Whenever a magnetic field is detected in this sample, it is transferred to the second sample. The second Targeted Component (TC) sample contained 6 systems at the start of the project, and has been build by isolating the previously selected SC systems that were reported to have at least one magnetic component. The purpose of this sample is to perform dense monitoring of each object in order to map the magnetic fields at the surface of the stars. Both samples will allow us to establish the magnetic properties (incidence, nature, geometry, strength) in close binary systems and compare them to single stars, already analysed in previous programs \citep[e.g. MiMeS,][]{wade13}.

The observational resources are mainly based on two high-resolution spectropolarimetric large programmes (LP) of $\sim$600 h and $\sim$150 h respectively allocated on ESPaDOnS (Canada-France-Hawaii Telescope) and Narval (T\'elescope Bernard Lyot, France). Data acquisition started in early 2013 and will end in late 2016. In order to better constrain the orbital parameters (inclination and flux ratio), or the stellar activity of the cool systems of our sample, we have started or proposed additional programs using complementary instrumentation: a PIONIER/VLTI program (PI: J.-B. Le Bouquin) is acquiring interferometric data of the magnetic BinaMIcS targets ; a CHANDRA proposal (PI: C. Argiroffi) has been submitted to analyse the X-ray activity of an eccentric pre-main-sequence system along the orbit, simultaneously with its magnetic activity ; photometric observations of cool systems will be obtained to improve the surface mapping of the components.

\section{First look analysis and preliminary results}

We will focus here only on the results we obtained so far in high-mass systems.



\begin{figure}[t]
\begin{minipage}[t]{.46\linewidth}
\centering
\includegraphics[width=5.5 cm]{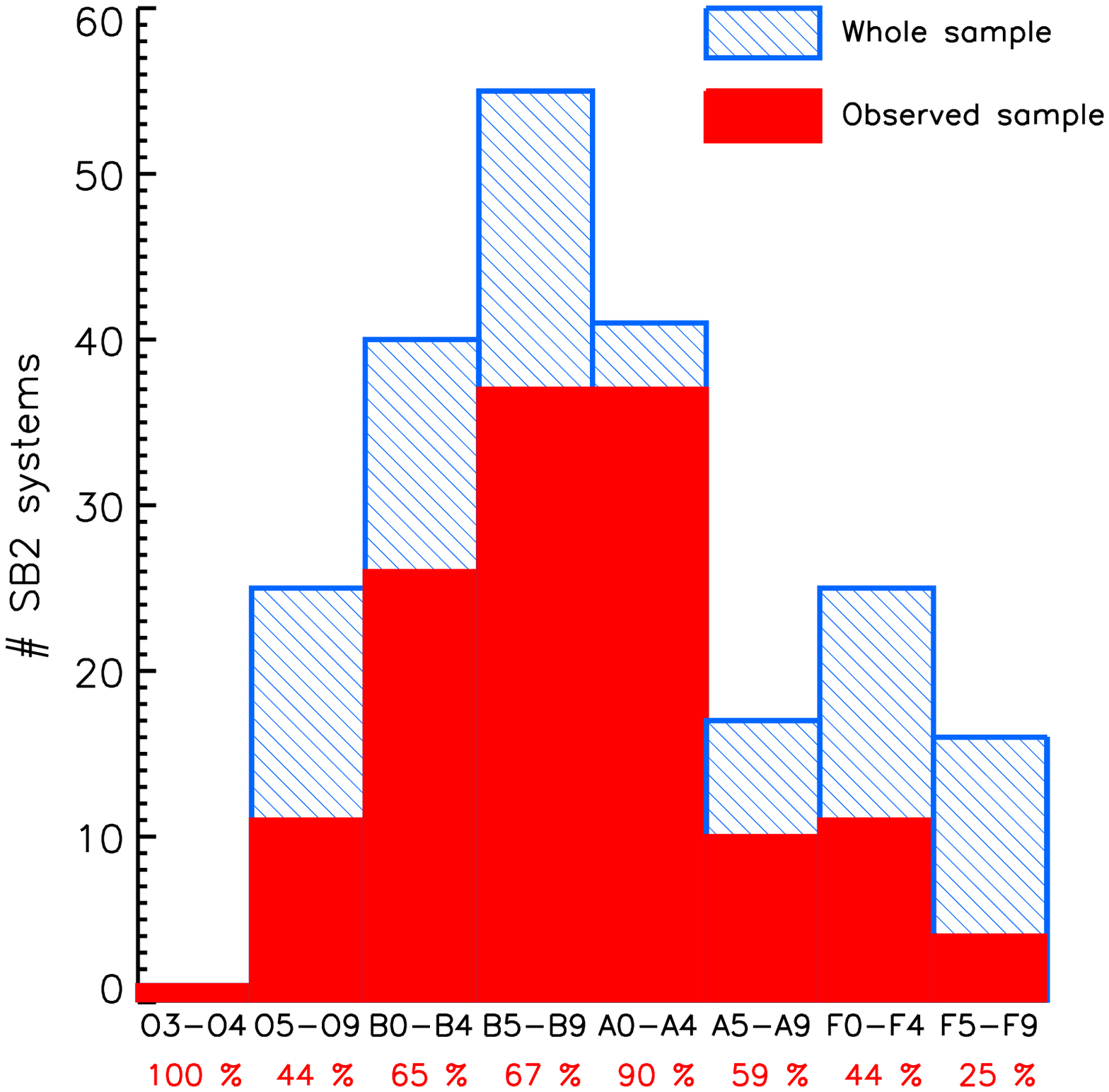} 
\caption{Observed (filled, red) and whole (dashed, blue) BinaMIcS samples as a function of the spectral type of the primary component. The fractions of targets already observed are indicated per spectral type at the bottom.}
\label{fig:spt}
\end{minipage} \hfill
\begin{minipage}[t]{.46\linewidth}
\centering
\includegraphics[width=5.5 cm]{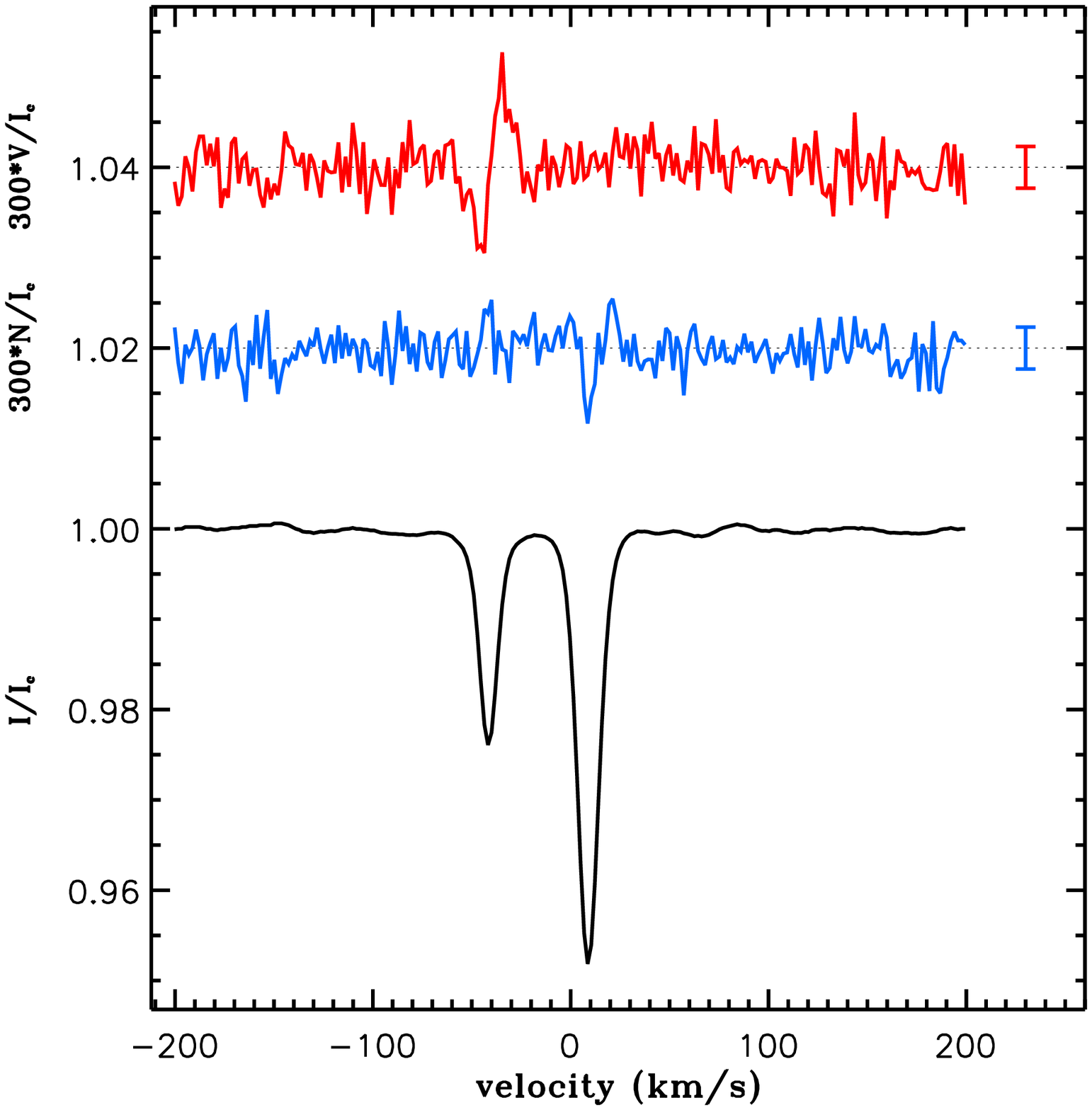} 
\caption{LSD $I$ (bottom), $V$ (top), $N$ (middle) profiles of the F4+F5 binary system HD~160922. The $V$ and $N$ profiles have been shifted in the Y-axis and amplified for display purpose. Note the Zeeman signature in the secondary (left) component only.}
\label{fig:hd160922}
\end{minipage}
\end{figure}

We have determined the exposure times of each SC target in order to reach a detection limit of 150 G or lower. These exposure times take into account the magnitude and binarity of the system, and the $v\sin i$ and spectral type of the stars. All these quantities can affect the amplitude of the detection of the magnetic field in the polarised spectra, hence the detection limit. The data have been reduced using the fully automatic Libre-Esprit reduction tool available at both telescopes. Libre-Esprit provides us with optimally reduced intensity $I$ and circularly polarised $V$ spectra. A null $N$ spectrum is also computed in a way such that polarisation is cancelled, allowing us to check for spurious polarisation in our data \citep{donati97}.

In order to increase the signal-to-noise ratio (SNR) of our data we applied the Least-Squares Deconvolution \citep[LSD,][]{donati97} technique to all our data. This technique provides us with the mean $I$, $V$, and $N$ profiles of the spectral lines with SNRs increased by a factor of 3 to 30 (depending on the spectral type) compared to SNRs in the individual spectral lines. This technique allows us to detect longitudinal magnetic fields as low as a few gauss with a reasonable exposure time.

Of the 220 SC targets, we have observed 151 of them. We have detected only one magnetic field so far, in the secondary component of HD 160922 \citep[Fig. \ref{fig:hd160922},][]{neiner13}. The spectral types of the primary and secondary components of HD 160922 are respectively F4 V and F5 V. A spectral type of F5 corresponds to the generally accepted limit below which the surface magnetic field is no longer likely to be fossil, and is of dynamo origin - while this limit still needs to be explored in more details. It is therefore not yet clear what the nature of the field detected in HD 160922 is. We will therefore assume that our number of fossil field detections, today, is between 0 and 1. If we include the 6 previously known TC targets with one magnetic component, and take into account that each system contains 2 stars, we now have gathered sensitive magnetic measurements in 314 high-mass stars belonging to close binary systems. 6 or 7 of them host a fossil field, leading to a fossil field incidence lower than 2\%. This is 3 to 5 times lower than what is observed in high-mass single stars. We can therefore conclude that fossil magnetic fields seem to be more absent in close binary systems than in single stars. This result is consistent with the work of Carrier et al. (2002) based on a much smaller sample, and only on A-type stars. Here our larger sample allows us to extend this result to higher masses, i.e. to all stars with radiative envelopes, and to quantify the differences between close binaries and single stars (Alecian et al. in prep.).

\section{Perspectives}

This very first result, while preliminary, is interesting to analyse in the context of magnetic impact on star formation. Recent theoretical developments have shown that weakly magnetised molecular cores can fragment, while the fragmentation is highly reduced in the case of strongly magnetised cores \citep[e.g.][]{commercon11}. In a very simple view, if initial conditions can be retained throughout the stellar formation process until the final stage of a binary star, then we would expect that single stars are more magnetised than binary stars, which is what we observe. However, detailed calculation of magnetised collapse tend to show that initial conditions are lost at the end of the first collapse (Masson et al., priv. comm.). Therefore, the properties of fossil fields in single and close binary systems might not come from the earliest stages of star formation, indicating that we still need to understand the discrepancy between single stars and close binary systems. Progress in our understanding of the magnetic impact in the later stages of star formation (from the second collapse to the Hayashi phase) appear crucial to unterstand the origin and evolution of fossil fields.

\bibliographystyle{iau307}
\bibliography{aleciane}

\begin{thebibliography}{}

\bibitem[\protect\astroncite{{Barker} \& {Ogilvie}}{2009}]{barker09}
{Barker}, A.~J. \& {Ogilvie}, G.~I. 2009,
\newblock {\em \mnras} 395, 2268

\bibitem[\protect\astroncite{{Bonnell} \& {Bate}}{1994}]{bonnell94}
{Bonnell}, I.~A. \& {Bate}, M.~R. 1994,
\newblock {\em \mnras} 271, 999

\bibitem[\protect\astroncite{{Borra} et~al.}{1982}]{borra82}
{Borra}, E.~F., {Landstreet}, J.~D., \& {Mestel}, L. 1982,
\newblock {\em \araa} 20, 191

\bibitem[\protect\astroncite{{Castor} et~al.}{1975}]{castor75}
{Castor}, J.~I., {Abbott}, D.~C., \& {Klein}, R.~I. 1975,
\newblock {\em \apj} 195, 157

\bibitem[\protect\astroncite{{C{\'e}bron} \& {Hollerbach}}{2014}]{cebron14}
{C{\'e}bron}, D. \& {Hollerbach}, R. 2014,
\newblock {\em \apjl} 789, L25

\bibitem[\protect\astroncite{{Commer{\c c}on} et~al.}{2011}]{commercon11}
{Commer{\c c}on}, B., {Hennebelle}, P., \& {Henning}, T. 2011,
\newblock {\em \apjl} 742, L9

\bibitem[\protect\astroncite{{Donati} \& {Landstreet}}{2009}]{donati09}
{Donati}, J.-F. \& {Landstreet}, J.~D. 2009,
\newblock {\em \araa} 47, 333

\bibitem[\protect\astroncite{{Donati} et~al.}{1997}]{donati97}
{Donati}, J.-F., {Semel}, M., {Carter}, B.~D., {Rees}, D.~E., \& {Collier
  Cameron}, A. 1997,
\newblock {\em \mnras} 291, 658

\bibitem[\protect\astroncite{{Gregory} et~al.}{2014}]{gregory14}
{Gregory}, S.~G., {Holzwarth}, V.~R., {Donati}, J.-F., {et~al.} 2014,
\newblock in {\em European Physical Journal Web of Conferences}, Vol.~64 of
  {\em European Physical Journal Web of Conferences}, p. 8009

\bibitem[\protect\astroncite{{Hut}}{1980}]{hut80}
{Hut}, P. 1980,
\newblock {\em \aap} 92, 167

\bibitem[\protect\astroncite{{Moss}}{2001}]{moss01}
{Moss}, D. 2001,
\newblock in G. {Mathys}, S.~K. {Solanki}, \& D.~T. {Wickramasinghe} (eds.),
  {\em Magnetic Fields Across the Hertzsprung-Russell Diagram}, Vol. 248 of
  {\em Astronomical Society of the Pacific Conference Series}, pp 305--+

\bibitem[\protect\astroncite{{Neiner} \& {Alecian}}{2013}]{neiner13}
{Neiner}, C. \& {Alecian}, E. 2013,
\newblock in {\em EAS Publications Series}, Vol.~64 of {\em EAS Publications
  Series}, pp 75--79

\bibitem[\protect\astroncite{{Ogilvie} \& {Lin}}{2007}]{ogilvie07}
{Ogilvie}, G.~I. \& {Lin}, D.~N.~C. 2007,
\newblock {\em \apj} 661, 1180

\bibitem[\protect\astroncite{{Parker}}{1960}]{parker60}
{Parker}, E.~N. 1960,
\newblock {\em \apj} 132, 175

\bibitem[\protect\astroncite{{Pourbaix} et~al.}{2009}]{pourbaix09}
{Pourbaix}, D., {Tokovinin}, A.~A., {Batten}, A.~H., {et~al.} 2009,
\newblock {\em VizieR Online Data Catalog} 1, 2020

\bibitem[\protect\astroncite{{Remus} et~al.}{2012}]{remus12}
{Remus}, F., {Mathis}, S., \& {Zahn}, J.-P. 2012,
\newblock {\em \aap} 544, A132

\bibitem[\protect\astroncite{{Roxburgh}}{1983}]{roxburgh83}
{Roxburgh}, I.~W. 1983,
\newblock in J.~O. {Stenflo} (ed.), {\em Solar and Stellar Magnetic Fields:
  Origins and Coronal Effects}, Vol. 102 of {\em IAU Symposium}, pp 449--459

\bibitem[\protect\astroncite{{Sana} \& {Evans}}{2011}]{sana11}
{Sana}, H. \& {Evans}, C.~J. 2011,
\newblock in C. {Neiner}, G. {Wade}, G. {Meynet}, \& G. {Peters} (eds.), {\em
  IAU Symposium}, Vol. 272 of {\em IAU Symposium}, pp 474--485

\bibitem[\protect\astroncite{{Taylor} et~al.}{2003}]{taylor03}
{Taylor}, S.~F., {McAlister}, H.~A., \& {Harvin}, J.~A. 2003,
\newblock in {\em American Astronomical Society Meeting Abstracts}, Vol.~35 of
  {\em Bulletin of the American Astronomical Society}, p. 1342

\bibitem[\protect\astroncite{{ud-Doula} et~al.}{2009}]{uddoula09}
{ud-Doula}, A., {Owocki}, S.~P., \& {Townsend}, R.~H.~D. 2009,
\newblock {\em \mnras} 392, 1022

\bibitem[\protect\astroncite{{Wade} et~al.}{2013}]{wade13}
{Wade}, G.~A., {Grunhut}, J., {Petit}, V., {et~al.} 2013,
\newblock in {\em Massive Stars: From alpha to Omega}

\end{thebibliography}

\begin{discussion}

\discuss{V. Khalack}{Can you say which part of your stars in the selected binary systems hvee slow axial rotation ($v\sin i < 20$~km.s$^{-1}$) ?}

\discuss{Authors}{Our spectra will be able to provide this information. We we have not retrieved it yet.}

\end{discussion}

\end{document}